\documentclass[preprint,aps,amsmath,showpacs,showkeys]{revtex4}
\usepackage{latexsym,amsmath,amssymb,amsbsy}
\usepackage[english]{babel}

\makeatother

\begin{document}

\title{A possible relation of the mass of the Universe\\ with the characteristic sizes of elementary particles}
\author{N.~A.~Miskinova}
\email{namisk@yandex.ru}
\author{B.~N.~Shvilkin}
\email{bshvilkin@yandex.ru}
\affiliation{Department of Physics, Moscow Technical University of Communications and Informatics,
\mbox{8a Aviamotornaya St., Moscow 111024, Russia}\\}
\affiliation{Department of Polymer and Crystal Physics, Faculty of Physics,
M. V. Lomonosov Moscow State University, Moscow 119991, Russia\vspace{1cm}}


\begin{abstract}
We show that the mass of the matter equal to the mass of the observable part of our Universe is reached at the Planck density in the volume which size is comparable with the nucleon size  and is close to the pion Compton wavelength.
\end{abstract}



\maketitle

The dimension analysis in physics often leads to the discovery of essentially important laws. Max Planck, for example, by the dimension analysis managed to introduce sizes, having dimensional lengths and time,  the so-called Planck  length and time \cite{r1}, the extreme  smallness of which have led to concept of discrete behavior of space and time. The Plank length ${\ell _{\rm{P}}} \sim {10^{ - 33}}\,\mbox{cm}$  defines the ``quantum" of dimensional distance, and the time ${\tau _{\rm{P}}} \sim {10^{ - 44}}\,\mbox{s}$ defines the time ``quantum". As a result from the dimension analysis the conclusion has been drawn on the necessity of building the quantum theory for discrete space-time \cite{r2} (see also \cite{r3} and the references therein).

Let's carry out the analysis of dimensional Planck values in application to the expanding Universe. At the beginning of expansion of the Universe the substance was in a so-called vacuum state \cite{r4}. The matter density was extremely great and was sustained constant. This density called Planck density is possible to express through Planck dimensional values:
\begin{equation}
\label{e1}
m_{\rm P} = \sqrt{\frac{\hbar c}{G}} \simeq 2.18 \times 10^{ - 5}\,\mbox{g},
\end{equation}
Planck mass, and
\begin{equation}
\label{e2}
\ell _{\rm P} = \frac{\hbar }{m_{\rm P}c} = \sqrt {\frac{\hbar G}{c^3}}  \simeq 1.62 \times 10^{- 33}\,\mbox{cm},
\end{equation}
Planck length, where $\hbar$ is Planck constant, $G$ is Newtonian gravitational constant, and
$c$ is speed of light in vacuum (here and below values of physical and astrophysical constants are taken from \cite{r5}).

The substance in a vacuum state was characterized by gravitational repulsion (negative pressure), and it caused the powerful initial push, which caused almost instant expansion of material, so-called inflation \cite{r4}.

At some instant of time the size of the Universe was insignificantly small, it was defined by the size of the Planck cell with volume $\ell _{\rm P}^3 \simeq 4.22 \times 10^{- 99}\,\mbox{cm}^3$. The density of matter in this cell was huge (see Eqs. (\ref{e1}) and (\ref{e2})):
\begin{equation}
\label{e3}
\rho _{\rm P} = \frac{m_{\rm P}}{\ell_{\rm P}^3} = \frac{c^5}{\hbar G^2} \simeq 5.15 \times 10^{93}\,\mbox{g}\, \mbox{cm}^{- 3}.
\end{equation}
And it was so despite of small mass of matter inside the Planck volume, of about $10^{-5}\,\mbox{g}$, and rather small number of nucleons ${m_{\rm{P}}}/{m_p} \simeq 1.30 \times {10^{19}}$, comparable with Loschmidt's number ${N_{\rm{L}}} \simeq 2.69 \times {10^{19}}$ (number of molecules in $1\,\mbox{sm}^3$ of ideal gas under normal conditions).

It is considered that the Planck density dividing quantum and classical space-time defines a state of matter which conditionally is taken over for the ``beginning" or the ``birth" of our Universe \cite{r4}.

While matter expanding, the volume occupied by it became more and more. Also the mass of substance, because of negative energy of gravitation, thus increased at almost constant density. So it proceeded until the vacuum substance through an insignificant instant had not turned in the quantum mode to the usual matter of the Universe. For this shortest time the Universe had swelled incredibly (see, for example, \cite{r4}).

Let's estimate, what mass of material with the Planck density can be concentrated in the sphere volume with radius equal to the Compton wavelength of the $\pi$ meson ${\mathchar'26\mkern-10mu\lambda _\pi } = \hbar /({m_\pi }c)$. As it is known, this size defines the range of nuclear forces and is close to the nucleon size (the rms charge radius of the proton ${r_p} \simeq 0.88 \times {10^{ - 13}}\,{\mbox{cm}}$ \cite{r5}) and to the radius of confinement (holding quarks and gluons inside hadrons) \cite{r6}. In the case of the $\pi ^0$ meson we have length and the corresponding volume
\begin{equation}
\label{e4}
{\mathchar'26\mkern-10mu\lambda _{{\pi ^0}}} \simeq 1.46 \times {10^{ - 13}}\,{\mbox{cm}},\quad {V_0} = \frac{{4\pi }}{3}\mathchar'26\mkern-10mu\lambda _{{\pi ^0}}^3 \simeq 3.12 \times {10^{ - 38}}\,{{\mbox{cm}}^3}.
\end{equation}
For the $\pi^+$ meson,
\begin{equation}
\label{e5}
{\mathchar'26\mkern-10mu\lambda _{{\pi ^ + }}} \simeq 1.41 \times {10^{ - 13}}\,\mbox{cm},\quad {V_ + } = \frac{{4\pi }}{3}\mathchar'26\mkern-10mu\lambda _{{\pi ^ + }}^3 \simeq 2.83 \times {10^{ - 38}}\,\mbox{cm}^3.
\end{equation}

The mass of matter in volume ${V_0}$ at the Planck density (\ref{e3}) is equal to
\begin{equation}
\label{e6}
M_0 = \rho _{\rm P}V_0 = \frac{4\pi}{3}m_{\rm P}\left( \frac{\mathchar'26\mkern-10mu\lambda _{\pi ^0}}{\ell _{\rm P}} \right)^3 \simeq 6.75 \times 10^{55}\,\mbox{g}.
\end{equation}
Similarly,
\begin{equation}
\label{e7}
M_+ = \rho _{\rm P}V_ + = \frac{4\pi}{3}m_{\rm P}\left(\frac{\mathchar'26\mkern-10mu\lambda _{\pi ^ +}}{\ell _{\rm P}}\right)^3 \simeq 6.10 \times 10^{55}\,\mbox{g}.
\end{equation}
Masses (\ref{e6}) and  (\ref{e7}) are equivalent to the total mass of approximately $4\times10^{79}$ nucleons.

The obtained values of masses of $M_0$ and $M_+$ appear to be of the same order of magnitude with the mass of the observable part of our Universe $M_U$. According to modern knowledge,
\begin{equation}
\label{e8}
M_U = \frac{4\pi}{3}\rho _c(ct_0)^3 \simeq 8.84 \times 10^{55}\,\mbox{g},
\end{equation}
where  $t_0 = 13.69(13) \times 10^9\,{\mbox{yr}}$ is the age of the Universe,  ${\rho _c} = 1.87835(19) \times {10^{ - 29}}{h^2}\,{\mbox{g}}~\mbox{cm}^{ - 3}$ is the critical density of the Universe, $h = 0.72(3)$ is the present day normalized Hubble expansion rate.

Comparison of the masses (\ref{e6})--(\ref{e8}) shows that difference between them is rather small:
\[
\frac{M_U - M_0}{M_U} \simeq 0.24,\quad \frac{M_U - M_ + }{M_U} \simeq 0.31,\quad \frac{M_0 - M_ +}{M_ +} \simeq 0.10.
\]

Let's define also the effective size $r_c$ by a relation  ${M_U} = (4\pi /3)\rho _{\rm P}r_c^3$, whence
\[
r_c = \left(\frac{3}{4\pi}\frac{M_U}{\rho _{\rm P}}\right)^{1/3} \simeq 1.60 \times {10^{ - 13}}\,\mbox{cm}.
\]
This size differs from the pion Compton wavelength a little (see Eqs. (\ref{e4}) and (\ref{e5})):
\[
\frac{r_c - \mathchar'26\mkern-10mu\lambda _{\pi ^0}}{\mathchar'26\mkern-10mu\lambda _{\pi ^0}} \simeq 0.09,\quad \frac{r_c - \mathchar'26\mkern-10mu\lambda _{\pi ^ +}}{\mathchar'26\mkern-10mu\lambda _{\pi ^ +}} \simeq 0.13.
\]

The presented data allow us to assume that the typical hadron size and the mass in the corresponding volume at the Planck density are the characteristic size and the critical mass at the Big Bang originated from the Planck length and mass at the birth of the Universe.

In conclusion, it is shown that the mass of a ``Planck nucleon" (a sphere with a radius of order of the size of the nucleon filled with matter of the Planck density) is close to the mass of the observable part of our Universe.

The authors thank Prof. A. V. Borisov for his interest in this work and for discussion of the results.

\end{document}